\documentclass[12pt]{article}%
\usepackage{amsmath,latexsym}
\usepackage{graphicx}
\usepackage{amsmath}
\usepackage{amsfonts}
\usepackage{amssymb}%
\begin{document}

\title{{Spherically symmetric wormholes of embedding class
one}}
   \author{
PETER K F KUHFITTIG*\\
\footnote{E-mail: kuhfitti@msoe.edu}
 \small Department of Mathematics, Milwaukee School of
Engineering,\\
\small Milwaukee, Wisconsin 53202-3109, USA}

\date{}
 \maketitle

\begin{abstract}\noindent
This paper generalizes an earlier result by
the author based on well-established embedding
theorems that connect the classical theory of
relativity to higher-dimensional spacetimes.
In particular, an $n$-dimensional Riemannian
space is said to be of class $m$ if $m+n$
is the lowest dimension of the flat space
in which the given space can be embedded.
To study traversable wormholes, we concentrate
on spacetimes that can be reduced to embedding
class one by a suitable transformation.  It
is subsequently shown that the extra degrees
of freedom from the embedding theory provide
the basis for a complete wormhole solution
in the sense of obtaining both the redshift
and shape functions. \\


\end{abstract}

\section{Introduction}\label{S:Introduction}

Wormholes are handles or tunnels in spacetime
that are able to connect widely separated
regions of our Universe and may even connect
entirely different universes \cite{MT88}.
Such wormholes can be described by the static
and spherically symmetric line element
\begin{equation}\label{E:wormhole}
ds^{2}=-e^{\nu(r)}dt^{2}+\frac{dr^2}{1-b(r)/r}
+r^{2}(d\theta^{2}+\text{sin}^{2}\theta\,
d\phi^{2}),
\end{equation}
using units in which $c=G=1$.  Here
$\nu=\nu(r)$ is called the \emph{redshift
function}, which must be everywhere finite
to avoid an event horizon.  The function
$b=b(r)$ is called the \emph{shape function}.
The spherical surface $r=r_0$ is the
\emph{throat} of the wormhole.  Here
$b(r)$ must satisfy the following
conditions: $b(r_0)=r_0$, $b(r)<r$
for $r>r_0$, and  $b'(r_0)<1$, called
the \emph{flare-out condition}.  We also
wish to assume that $b'(r)>0$ due to the
field equation $8\pi\rho(r)=b'(r)/r^2$,
where $\rho$ is the energy density,
normally considered positive.  The
flare-out condition can only be
satisfied by violating the null energy
condition, discussed in Sec.
\ref{S:other}.  For a Morris-Thorne
wormhole, this violation requires the
use of ``exotic matter."

The discussion in Ref. \cite{MT88} is
based on the following strategy: specify
the geometric conditions required for a
traversable wormhole and then either
manufacture or do a search for
matter or fields that can produce the
desired energy-momentum tensor.  The
main goal of this paper is to reverse
this strategy by showing that the
conditions discussed are sufficient for
producing a complete solution, i.e., for
producing both the redshift and shape
functions.  The approach in this paper
differs significantly from that in Ref.
\cite{pK18}, which discusses charged
wormholes admitting a one-parameter
group of conformal motions, together
with a new model to explain the flat
galactic rotation curves without the
need for dark matter.

\section{The embedding}\label{S:embedding}
Unlike Ref. \cite{pK18}, the conditions
discussed in this paper are derived directly
from the assumption that the spacetime is
of embedding class one.  Here we need
to recall that an $n$-dimensional
Riemannian space is said to be of embedding
class $m$ if $m+n$ is the lowest dimension
of the flat space in which the given space
can be embedded \cite{MG17, M1, M2,
M3, M4, sM}.  It is well known that the
exterior Schwarzschild solution is a
Riemannian space of embedding class two.
Following Ref. \cite{MG17}, we start
with the static and spherically symmetric
line element
\begin{equation}\label{E:line1}
ds^{2}=e^{\nu(r)}dt^{2}-e^{\lambda(r)}dr^{2}
-r^{2}\left(d\theta^{2}+\sin^{2}\theta \,d\phi^{2}
\right).
\end{equation}
(For physical reasons, it is generally assumed
that $\nu(r)$ is finite and that
$\text{lim}_{r\rightarrow \infty}\nu(r)=0$.)
It is shown in Ref. \cite{MG17} that this metric
of class two can be reduced to a metric of class
one and can therefore be embedded in a
five-dimensional flat spacetime.
The following transformation can accomplish this
reduction:
 $z^1=r\,\text{sin}\,\theta\,\text{cos}\,\phi$, $z^2=
 r\,\text{sin}\,\theta\,\text{sin}\,\phi$,
 $z^3=r\,\text{cos}\,\theta$, $z^4=\sqrt{K}\,e^{\frac{\nu}{2}}
 \,\text{cosh}{\frac{t}{\sqrt{K}}}$, and $z^5=\sqrt{K}
 \,e^{\frac{\nu}{2}}\,\text{sinh}{\frac{t}{\sqrt{K}}}$.
 The result is \cite{MG17}
\begin{equation}\label{E:line3}
ds^{2}=e^{\nu}dt^{2}-\left(\,1+\frac{K\,e^{\nu}}{4}\,
{\nu'}^2\,\right)\,dr^{2}-r^{2}\left(d\theta^{2}
+\sin^{2}\theta\, d\phi^{2} \right).
\end{equation}
Metric (\ref{E:line3}) is therefore equivalent to
metric (\ref{E:line1}) if
\begin{equation}\label{E:lambda1}
e^{\lambda}=1+\frac{K\,e^{\nu}}{4}\,{\nu'}^2,
\end{equation}
where $K>0$ is a free parameter.  The condition
is equivalent to the following condition due to
Karmarkar \cite{kK48}:
\begin{equation}\label{E:Kar}
   R_{1414}=\frac{R_{1212}R_{3434}-R_{1224}R_{1334}}
   {R_{2323}},\quad R_{2323}\neq 0.
\end{equation}
(See Ref. \cite{pB16} for further details.)  So
while Eq. (\ref{E:Kar}) provides the justification
for the above embedding process, Eq.
(\ref{E:line3}) yields a useful mathematical
model, helped by the free parameter $K$.
Moreover, this model is consistent with the
induced-matter theory in Ref. \cite{pW92},
discussed further in Sec. \ref{S:other}.


\section{The solution}
To produce the desired wormhole solution, we
prefer the opposite signature in line element
(\ref{E:line1}) in order to be consistent with
line element (\ref{E:wormhole}):
\begin{equation}\label{E:line4}
ds^{2}=-e^{\nu(r)}dt^{2}+e^{\lambda(r)}dr^{2}
+r^{2}\left(d\theta^{2}+\sin^{2}\theta \,d\phi^{2}
\right).
\end{equation}
Most importantly, no additional assumptions will
be made regarding $\nu=\nu(r)$.

The shape function $b=b(r)$ has to incorporate
Eq. (\ref{E:lambda1}) on account of the
embedding.  An entire class of such shape
functions can be readily obtained by inspection:
 \begin{equation}\label{E:shape3}
   b(r)=r\left(1-\frac{1}
   {1+\frac{1}{4}Ke^{\nu(r)}[\nu'(r)]^2}
   \right)+\frac{r^n/r_0^{n-1}}
   {1+\frac{1}{4}Ke^{\nu(r_0)}[\nu'(r_0)]^2}.
\end{equation}
Observe that $b(r_0)=r_0$ for all $n$.  Our
main task is to show that these shape
functions satisfy all the other requirements
for shape functions.  That is the topic
of the next section.


\section{The condition $0<b'(r_0)<1$}
     \label{S:condition}
As noted in the Introduction, we need to
examine the condition $0<b'(r_0)<1$.  So
we start with
\begin{multline}\label{E:bprime}
   b'(r_0)=1-\frac{1}
   {1+\frac{1}{4}Ke^{\nu(r_0)}[\nu'(r_0)]^2}
   +\frac{n}
   {1+\frac{1}{4}Ke^{\nu(r_0)}[\nu'(r_0)]^2}\\
   +r_0\left
   (1+\frac{1}{4}Ke^{\nu(r_0)}[\nu'(r_0)]^2
   \right)^{-2}\frac{1}{4}Ke^{\nu(r_0)}
   \left(2\nu'(r_0)\nu''(r_0)
       +[\nu'(r_0]^3\right).
\end{multline}
To simplify the analysis, let us introduce
the following notations:
\begin{equation}\label{E:A}
   A=\frac{1}{4}e^{\nu(r_0)}[\nu'(r_0)]^2
\end{equation}
and
\begin{equation}\label{E:Omega}
   \Omega=
   e^{\nu(r_0)}\left(2\nu'(r_0)\nu''(r_0)
       +[\nu'(r_0]^3\right).
\end{equation}
In view of Eq. (\ref{E:bprime}), the condition
$0<b'(r_0)<1$ now yields
\begin{equation}\label{E:master}
   \frac{\left(\frac{1-n}{1+AK}-1\right)
   (1+AK)^2}{\frac{1}{4}r_0K}<\Omega<
   \frac{(1-n)(1+AK)}{\frac{1}{4}r_0K}.
\end{equation}
In the trivial case $\nu'(r_0)=0$, the
condition $0<b'(r_0)<1$ is satisfied provided
that $0<n<1$.  Accordingly, we need to
concentrate  on the nontrivial case
$\nu'(r_0)\neq 0$;  as a result, $A$ is
positive but $\Omega$ can be positive or
negative.  So the right-hand side of
Inequality (\ref{E:master}) is equivalent
to the flare-out condition $b'(r)<1$ at
or near the throat, while the left-hand
side is equivalent to $b'(r_0)>0$.  We
will consider the two cases separately.

\subsection{The condition $b'(r_0)<1$}
 To analyze the flare-out condition, we need
 to consider the two cases, $\Omega>0$ and
 $\Omega<0$.  To this end, let us rewrite the
 right-hand side of Inequality (\ref{E:master})
 as follows:
 \begin{equation}\label{E:K}
    K\left(\frac{1}{4}r_0\Omega-A(1-n)\right)
       <1-n.
 \end{equation}
 If $\Omega>0$, then $n$ must be less than 1
 to keep $K$ positive.  It also becomes
 apparent that $r_0$ is another free parameter.
 So we can choose $r_0$ large enough so that
 \begin{equation*}\label{E:free1}
    \frac{1}{4}r_0\Omega>A(1-n).
 \end{equation*}
 As a result,
\begin{equation}\label{E:K1a}
   K<\frac{1-n}{\frac{1}{4}r_0\Omega-A(1-n)},
   \quad n<1.
\end{equation}

In Inequality (\ref{E:K}), if $\Omega<0$, then
we must have $n>1$ to keep $K$ positive.  This
time we need to choose $r_0$ sufficiently large
so that
\begin{equation*}\label{E:free2}
   \frac{1}{4}r_0|\Omega|>-A(1-n).
\end{equation*}
The result is
\begin{equation}\label{E:K1b}
   K>\frac{1-n}{\frac{1}{4}r_0\Omega-A(1-n)},
   \quad n>1.
\end{equation}

It should be noted that Conditions (\ref{E:K1a})
and (\ref{E:K1b}) for the free parameter $K$
can always be met by increasing the throat size
of the wormhole.  Observe also that $n\neq 1$.

\subsection{The condition $b'(r_0)>0$}

The left-hand side of Inequality (\ref{E:master})
is more difficult to analyze since, after
simplifying, we get the quadratic inequality
\begin{equation}\label{E:quadratic}
   A^2K^2+K\left(\frac{1}{4}r_0\Omega
   +A+nA\right)+n>0.
\end{equation}
Once again, we need to consider the two cases
$\Omega>0$, $n<1$, and $\Omega<0$, $n>1$.

$\boldsymbol{\Omega >0, n<1}:$\, If
$\Omega>0$ and $0\le n<1$, Inequality
(\ref{E:quadratic}) is automatically satisfied
and we have $b'(r_0)>0$.

If $n<0$, we first need to solve the quadratic
inequality to obtain
\begin{equation}\label{E:K2}
   K<
   \frac{-\left(\frac{1}{4}r_0\Omega +A +nA\right)
   -\sqrt{\left(\frac{1}{4}r_0\Omega +A +nA\right)^2
   -4A^2n}}{2A^2}
\end{equation}
or
\begin{equation}\label{E:K3}
   K>
   \frac{-\left(\frac{1}{4}r_0\Omega +A +nA\right)
   +\sqrt{\left(\frac{1}{4}r_0\Omega +A +nA\right)^2
   -4A^2n}}{2A^2}.
\end{equation}
Algebraically, the solution is valid for both
$\Omega>0$ and $\Omega<0$.  Because of the ``or,"
only one of the inequalities is actually needed.
(Since $K$ has to be positive,  the first
inequality is unphysical anyway.)  For the second
inequality, $K>0$ since $n<0$.  We conclude that
for the case $\Omega>0$, $n<0$, the parameter $K$
must satisfy the following inequality:
\begin{equation}\label{E:K4}
   \frac{-\left(\frac{1}{4}r_0\Omega +A +nA\right)
   +\sqrt{\left(\frac{1}{4}r_0\Omega +A +nA\right)^2
   -4A^2n}}{2A^2}\\<K<
   \frac{1-n}{\frac{1}{4}r_0\Omega-A(1-n)},
      \quad n<0,
\end{equation}
referring back to Inequality (\ref{E:K1a}).  So
if $\Omega >0$ and $n<0$, then $K$ must lie
between two positive values.  We therefore have
a solution for the case  $\Omega >0, n<1$.

$\boldsymbol{\Omega <0, n>1}:$\, For the case
$\Omega<0$, $n>1$, the real difficulty is that
solutions (\ref{E:K2}) and (\ref{E:K3}) may
not be real.  To avoid this problem, let us
choose the free parameter $r_0$ sufficiently
large to start with, i.e., choose $r_0$ so that
$\frac{1}{4}r_0\Omega=-bA$ for some
sufficiently large positive constant $b$ to
obtain
\begin{equation}\label{E:real}
   (-bA+A+nA)^2-4nA^2>0,
\end{equation}
thereby resulting in a real solution.
Consequently, Inequality (\ref{E:K2}) yields
\begin{equation}\label{E:K5}
   K<\frac{-(-b+1+n)-\sqrt{(-b+1+n)^2-4n}}
   {2A}
\end{equation}
while Inequality (\ref{E:K1b}) gives
\begin{equation}\label{E:K6}
   K>\frac{1-n}{\frac{1}{4}r_0\Omega-A(1-n)}=
   \frac{2(1-n)/(-b-1+n)}{2A}.
\end{equation}
(Inequality (\ref{E:K3}) is not needed.)

The significance of the conditions on $K$ can
best seen graphically.  Fig. 1 shows that for
any fixed $n$,
\begin{figure}[tbp]
\begin{center}
\includegraphics[width=0.8\textwidth]{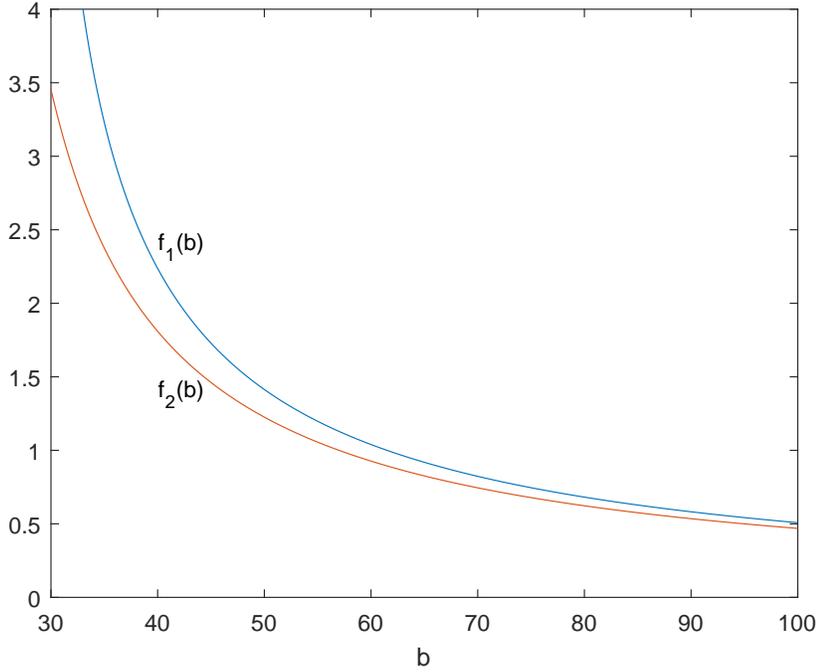}
\end{center}
\caption{Plots showing $f_1(b)$ and $f_2(b)$.}
\end{figure}
\begin{equation}
   f_1(b)=-(-b+1+n)-\sqrt{(-b+1+n)^2-4n}>f_2(b)\
   =\frac{2(1-n)}{-b-1+n},
\end{equation}
referring to Inequalities (\ref{E:K5}) and
(\ref{E:K6}).  So once again, $K$ must lie between
two positive values.  We therefore have a solution
for the case $\Omega<0$, $n>1$, as well.

\section{Other conditions}\label{S:other}

Having shown that the flare-out condition
$b'(r_0)<1$ has been met, let us return to the
violation of the null energy condition (NEC),
which states that for the energy-momentum
tensor  $T_{\alpha\beta}$,
\[
   T_{\alpha\beta}\mu^{\alpha}\mu^{\beta}\ge 0
\]
for all null vectors.  So given the radial
outgoing null vector $(1,1,0,0)$, we have that
$\rho(r_0) +p_r(r_0)<0$ whenever the condition
is violated. By Ref. \cite{MT88}, this
violation is equivalent to the condition
\begin{equation}
   \frac{b'(r_0)-b(r_0)/r_0}{2[b(r_0)]^2}<0,
\end{equation}
which holds whenever $b'(r)<1$ at or near
the throat.  As noted in the Introduction,
for a Morris-Thorne wormhole, the violation
of the NEC requires the use of ``exotic
matter," since ordinary matter normally
satisfies the NEC.  We have seen, however,
that the shape functions and subsequent
flare-out conditions were obtained from
the embedding theory, which may be viewed
as part of the induced-matter theory
\cite {pW92} in the following sense:
according to Ref. \cite{pW15}, the field
equations for the five-dimensional flat
embedding space yield the Einstein field
equations in four dimensions
\emph{containing matter}.  The
induced-matter theory therefore implies
that the matter in our Universe actually
comes from geometry and this may very well
include exotic matter.  So while exotic
matter cannot be avoided, it may be less
problematical in the present context.

Our final observation concerns asymptotic
flatness.  Since $\nu(r)\rightarrow
0$ as $r\rightarrow\infty$, we also have
$\text{lim}_{r\rightarrow \infty}\nu'(r)=0$.
So if $n<1$, we see from Eq. (\ref{E:shape3})
that $b(r)/r\rightarrow 0$ (in addition to
$e^{\nu(r)}\rightarrow 1$), resulting in an
asymptotically flat spacetime.

Unfortunately, this conclusion does not hold
for $n>1$.  So the wormhole spacetime has
to be cut off at some $r=a$ and joined to
an external Schwarzschild spacetime
\begin{equation}
ds^{2}=-\left(1-\frac{2M}{r}\right)dt^{2}
+\frac{dr^2}{1-2M/r}
+r^{2}(d\theta^{2}+\text{sin}^{2}\theta\,
d\phi^{2})
\end{equation}
in the usual way.  From $e^{\nu(a)}=
1-2M/a$, we have $2M=a\left(1-e^{\nu(a)}
\right)$.  But $2M=b(a)$; so the cut-off
at $r=a$ is implicitly determined by the
equation $b(a)=\left(1-e^{\nu(a)}\right)$,
provided, of course, that such a solution
exists.

\section{Conclusions}
An $n$-dimensional Riemannian space is
said to be of embedding class $m$ if
$m+n$ is the lowest dimension of the flat
space in which the given space can be
embedded.  Following Ref. \cite{MG17}, we
assume a spherically symmetric metric of
embedding class two that can be reduced
to class one by a suitable transformation.

These ideas were applied toward obtaining
a complete wormhole solution without the
usual engineering considerations, i.e.,
without being required to find or to
manufacture matter or fields that produce
the desired energy-momentum tensor.  The
free parameters $K$ and $r_0$ provided the
extra degrees of freedom to obtain both
the redshift and shape functions from the
embedding theory and may even account for
exotic matter.

\end{document}